# A Distributed Multi Agents Based Platform for High Performance Computing Infrastructures

Chairi Kiourt, Dimitris Kalles

School of Sciences and Technology, Hellenic Open University, Patras, Greece
{chairik, kalles}@eap.gr

**Abstract.** This work introduces a novel, modular, layered web based platform for managing machine learning experiments on grid-based High Performance Computing infrastructures. The coupling of the communication services offered by the grid, with an administration layer and conventional web server programming, via a data synchronization utility, leads to the straightforward development of a web-based user interface that allows the monitoring and managing of diverse online distributed computing applications. It also introduces an experiment generation and monitoring tool particularly suitable for investigating machine learning in game playing. The platform is demonstrated with experiments for two different games.

**Keywords:** Distributed Computing, MABS platform, MAS, Machine Learning.

## 1 Introduction

In almost all subareas of computing, there are great needs for powerful computational infrastructure; examples of such fields are data mining, 3D modeling, graphics rendering, large scale simulations, amongst others. Researchers from many scientific areas of Artificial Intelligence (AI) argue that, a combination of Multi Agent Based Simulations (MABS) [1] and High Performance Computing (HPC) systems are required for experiments with large agent numbers [2], especially so in Social Learning (SL) experiments with Multi Agent Systems (MAS) [1]. Modeling and monitoring of such environments requires powerful simulation tools [3]. Our case study is about large-scale Multi-Agent Based Simulation (MABS) systems, which facilitate the study of complex Social Environments (SE) composed of special agent-characters and other objects [1][4].

In most cases, large scale experiments and MABS systems require high computing resources [5], usually available via Grid Infrastructures (GI) [6]. These have driven the development of Agent Based Modeling (ABM) [1], [3] platforms, with promising potential but with limitations as well. Some of the most well-known agent based platforms are Repast HPC [7] and Flame [8]. The most common limitation of those platforms is that experiments run on single Worker Nodes (WN) in Computer Clusters. The distribution of the experiment has to be programmed by the users, including external distribution libraries. This limitation is overcome by our platform.

The main contribution of the platform presented in this paper is the experiment distribution ability to the GIs for faster execution of the experiment and for improved use of the resources and the subsequent streamlining of collecting the experimental results. Our platform can be thought of as filling the middle ground between two powerful extremes; the very complicated middleware's of GIs (such as *gLite* [9]) and workflow systems [10]; both demand a substantial learning curve whereas we propose a platform where some generic functionality is sacrificed for the benefit of lowering the technical barrier to entry. Furthermore, our platform parallelizes the experiment by splitting it into sub-experiments running autonomously in available WNs of GIs. It also caters to users with little expertise on GIs, as most functions are automated, generating log files, for each experiment, which may be subsequently analyzed. Additionally, the web-based implementation of the platform provides useful availability and portability, with a friendly Graphical User Interface (GUI), which is an important advantage compared to other platforms.

The rest of this paper is organized as follows: the next section presents a brief overview of background on distributed computing and on systems using a layered MABS platform structure. The third section introduces our proposed enhancement of the platform at the grid infrastructure level, which issued as the synchronization tool between the user interface and the grid-based computational processes. The fourth section introduces a tool for developing large-scale Machine Learning game playing experiments. The final section discusses and concludes the paper; it also offers some important future directions.

## 2    Background Knowledge

High Performance Computing software and hardware architectures have recently emerged to support large scale experiments, where desktop or organization-wide resources are considered limited.

### 2.1    Usage of grid infrastructures and MABS platforms

Nearly all scientists can use the Grid [6], on topics ranging from computer science and engineering to experimental sciences, operations research, and environmental simulation. Recently [13][14], the class of users has expanded to cover nations, states, corporations and private consumers.

While computer social simulation began to be studied widely in the last 20 years [6], in almost all projects the lack of computer resources was evident, particularly in modeling and understanding social processes [6]. The modeling of those systems was presented as an Agent Based Modeling system, creating autonomous elements on simulation models to facilitate the study of social environments. In the direction of more effectively studying such systems, visualization tools have been developed for creating and managing experiments giving rise to the term MABS.

## 2.2 Multi Agent Based Systems Platform

Our platform focuses on users working on large scale Machine Learning (ML) experiments with the need of computational infrastructures. We have tested it with social game playing experiments [11] [4], over many iterations. We have primarily used it in RLGame [11], where we research learning in games; however, we also demonstrate it on the Rock Scissors Paper (RSP) [16] game, as a tutoring example for fellow researchers (see the Appendix). The common elements of RLGame and RPS are they both are zero-sum games, which may demonstrate a playing behavior that may be volatile and can be used as the basis of social event type [4] experiments (large tournaments). We have made the source code of the RPS game in our platform available to all fellow researchers, to showcase the platform functionality and to assist fellow researchers in developing their own experiments with their own games.

## 2.3 RLGame

RLGame [12] is a strategy board game, which consists of two agents and their pawns and is played on an n x n square board. Two a x a square bases are on opposite board corners; these are initially populated by $\beta$ pawns for each player. The goal for each agent is to move a pawn into the opponent's base or to force all opponent pawns out of the board. The workbench, RLGame, was initially presented as a Competition extreme for studying multi agent systems via its tournament version, RLGTournament [4][25], implementing a Round-Robin (RR) scheme to pair participants against each other.

The learning mechanism of each player is based on approximating its value function with a artificial neural network and the Temporal Difference (TD) [15] learning mechanism of each player. As input layer nodes we use the board positions for the next possible move, plus some flags on overall board coverage. The hidden layer consists of half as many hidden nodes. A single node in the output layer, denotes the extent of the expectation to win when one starts from a specific game-board configuration and then makes a specific move.

## 2.4 Rock Scissors Paper

Rock Scissors Paper [16] is a zero-sum game widely use in Artificial Intelligence as a tool for various studies. In this game, each agent chooses, without the other agent's knowledge, one of three possible moves: Rock, Scissors or Paper. After the moves have been selected, they are revealed, and the winner is determined according to simple set of rules that have a playful interpretation: Rock "breaks" Scissors, Scissors "cuts" Paper, and Paper "covers" Rock. If the same move is played by both agents, the game is declared a draw.

The Rock Scissors Paper game is described in [23], which involves the agent repeatedly playing Rock Scissors Paper against an exploitable opponent which also implements the Active-LZ [23] algorithm combined with a Lempel-Ziv algorithm based on prediction scheme with dynamic programming for control to produce an agent that is provably asymptotically optimal if the environment is n-Markov. It also has been the subject of two computer tournaments for several years [24].

## 2.5 Related Work

A recently built agent based coalition formation model to study agent trust [17] resulted in a recommendation to adopt a high performance agent based computing platform [2] for exploring social simulation large-scale scenarios; that work adopted existing systems: a MABS platform [2] and Repast HPC [18]. Decraene *et al.* [19] developed an agent based system with a strong HPC component due to its high complexity; numerous iterations, diverse models, and the need for experiment design support and analysis of simulations led them to the above combination. The simulation models were represented using XML files [19]. In Pandora, an open-source framework for designing MAS experiments on HPC systems, one witnessed the deployment of a high-performance agent-based modeling framework on clouds [20]. Overall, the net result seems to be inconclusive, as solutions seem to be highly dependent on budget constraints and case specifics, thus compromising generalization efforts. Blanchart *et al.* present a novel architecture MABS platform on a cluster or a grid to generate data for scientific use [21]. They recommended that MABS and HPC infrastructures have to be coupled, especially in cases of social simulations with need of computational resources [21]. Their system also uses XML file to represent all possible simulations, while cluster access and management was implemented via a web based GUI [21].

As clearly indicated above, large scale multi agent systems are becoming an inevitable part of the HPC environment, including social simulation experiments, and though many individual differences are present, the underlying architectural concept of putting HPC and MAS systems together still thrives [22].

## 3  Platform Structure

The platform is separated into three layers, each of which consists of several objects and sub-objects. Figure 1a, illustrates the structure of the communication between layers and the objects, which is based on XML files. A bottom layer contains the principal objects of an experiment: agents, the game, scheduling algorithms, etc. The middle layer is the MABS platform, a system which not only routes communication between the Objects and the Monitoring layers, but also manages the process structure of the experiment. The top layer is the GUI from which user interfaces with experiments and GIs.

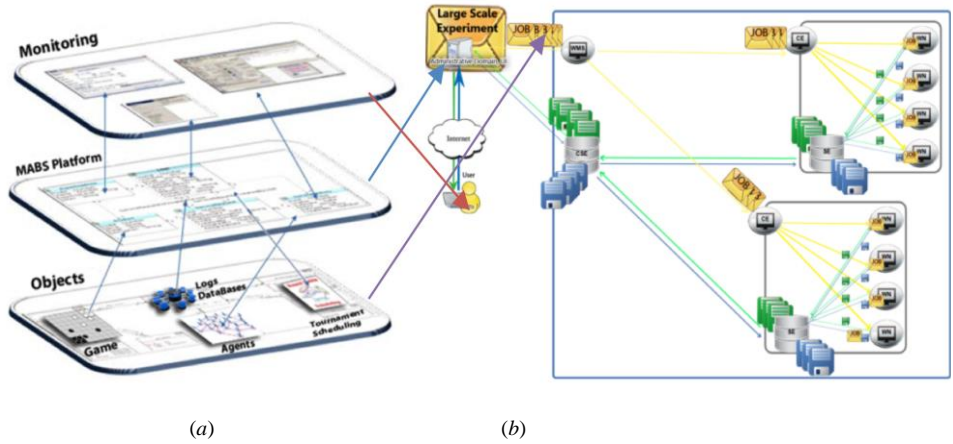

(a) (b)
**Fig. 1.** (a) Platform layers, (b) Large scale experiment Segmentation in HPC

After transferring the experimental data to the UI of the Grid Infrastructures, the experiment begins with its segmentation according to the RR algorithm. The large folder of Fig. 1a, represents the MAS experiment and its input and output data. After the segmentation of the experiment, the system of the platform begins to transfer the necessary data of each match–job to the Storage Element (SE), in parallel, (most of the blue Floppy Disks (FP) on the left in Fig. 1b). During data transfer the system executes the autonomous jobs and finds the available Clusters of the HPC Infrastructures through the Worker Management System (WMS). Thereafter, the Computer Element (CE) of each Cluster searches for free WNs while transferring the necessary data of the experiment from the SE to its own Storage Element. As soon as the available WN is found and after the influx of necessary data from the SE to the WN is completed, the job starts. After the job starts a unique Job ID is created and made available for the UI and for the web-based GUI for monitoring the status of the job. After the end of the job, by reverse flow the system of the platform brings the result of the experiment and the newly generated data to the UI (green FD) in order to be ready for the next round of the RR. This process is same for all the matches of a Social Event. It can also simultaneously manage hundreds of jobs for faster completion of a large-scale MAS experiment, which cannot run on a simple computer or on a single WN, and thus better utilize the Grid Resources.

The main architecture of the platform is based on the development of Round Robin (RR) algorithm based on social experiments. Put simply, multi agent experiments are decomposed in sub-experiment (matches) based on RR algorithm running in parallel, with many repetitions per match. Each match is an autonomous Grid Job [6] which is submitted automatically from the platform to any available WN of the GIs, by carrying all necessary input and output data to and from WN, as it is presented in Fig. 1b.

These MABS platform services and GIs services can be managed through a new built web based JAVA applet, which is implemented as a comprehensive SSH communication protocol. All the attributes of the platform is supported through a user-friendly GUI. This application was built over the existing infrastructures, with a view to safer and faster access to Grid and our MABS services.

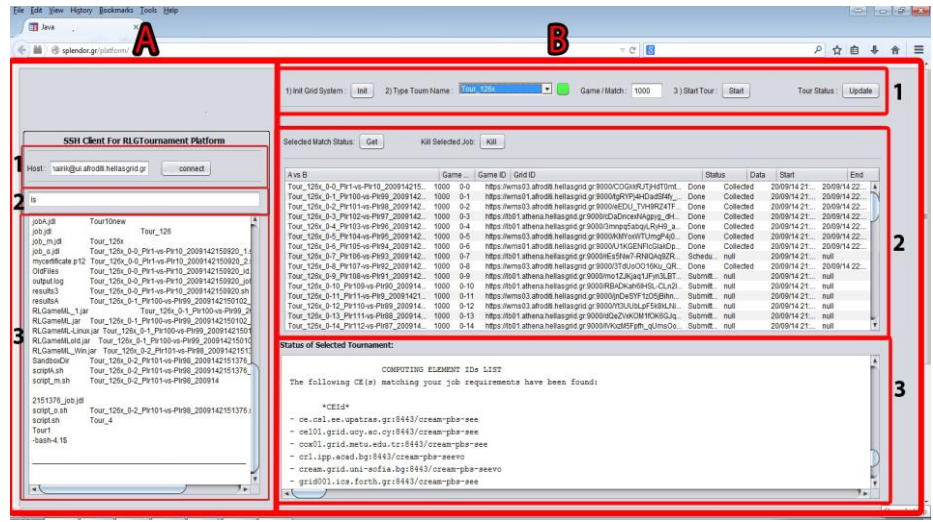

**Fig. 2.** Experiments Management GUI

Fig. 2 presents the GUI of the application, which is separated into double column panels, A and B, where each one consists of some sub-panels. Panel A is an implementation of a full comprehensive safe SSH terminal, which allows the communication between the grid User Interface (UI) and user terminal. The B panel is the main controller of the MABS platform, from where all the available experiments can be managed. The entire status information of the experiments and their jobs is stored in XML file, which represents the full map of the experiments and is shown in the GUI. It also improves the automatical re-submission of the experiments, which were stopped because of various reasons.

Every experiment requires some data pre-configurations as inputs. To facilitate the pre-configuration procedures, a tool, which is an important key for creating agents with different characters in common social event, is developed. This tool is shown in the Fig. 3, the process of which starts from the leftmost frame. The first step of the tool usage is to create the experiment and the next step is to add the agents and to set their character configuration. The second step can be repeated for additional agents with different characters. This tool ends its operation by exporting a folder with the pre-configuration files, according to experiment requirements, which can be transferred in all possible experiment execution resources, in this case Grid HPC Infrastructures.

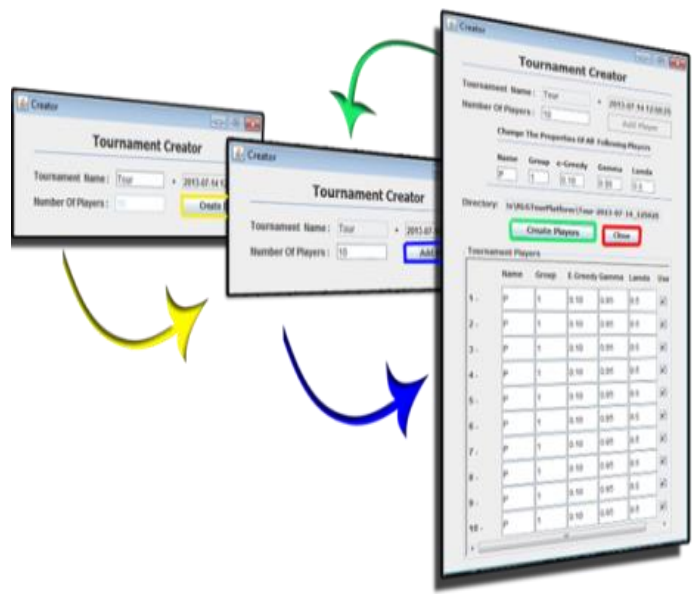

**Fig. 3.** Social Event Developer Tool

At this point, it should be emphasized that this platform does not override any of the powerful services (middleware's) of Grid Infrastructures, but it is added as an additional powerful tool with a focus on not only large scale Machine Learning experiments, but also on many others.

## 4      Experimental Demonstration

To enhance the reliability, the quality and the usage of the presented platform, as well as for debugging purposes, two independent experiments were developed with the same pre-configurations; these are TourA and TourB. Experiments using the RLGame were initiated with 126 agent in a RR tournament with 100 games per match. Each agent played 125 matches against all other agents, accounting for 12,500 games per agent and totaling 1,575,000 games per experiment.

In a personal computer with 2 cores and 4 GB RAM, a match of 100 games between two inexperienced agents (i.e. at the start of the experimental session, before learning experience starts to accumulate) takes about 1 minute to conclude. Extrapolating that to a full experiment, as set out in the previous paragraph, one arrives at an estimate of 26,000 hours to finish.

By exploiting the distributed infrastructure available via our platform, we spent about 24 days per experiment, roughly corresponding to about $1/50^{th}$ of the elapsed time of the sequential experiment, due to the substantial involvement of many computer systems in the grid infrastructure. Moreover, the automatic handling of problems, exceptions, etc. by our system and the subsequent facilitation of re-scheduling the problematic experiments meant that a further substantial time saving was realized as regards the time that a user spends to monitor the HPC infrastructure. For the sake of completeness, Fig. 4 presents some usage statistics on the grid nodes utilized by the experiment.

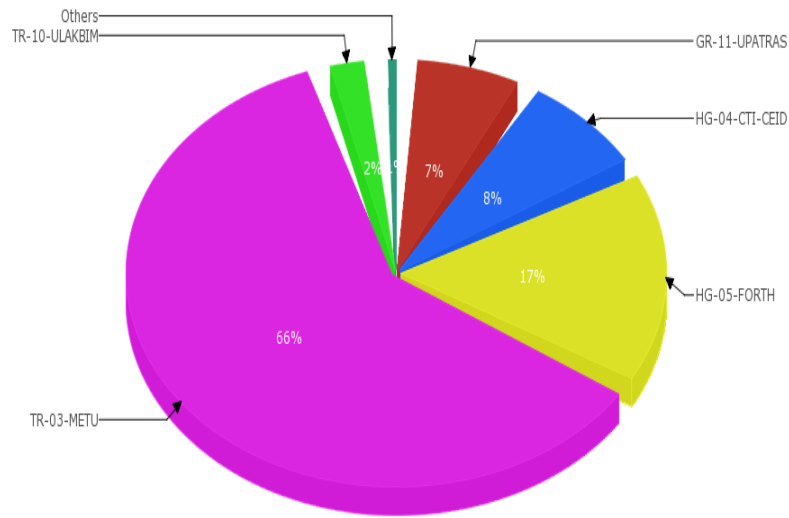

**Fig. 4.** Grid Infrastructure Side Usage

## 5 Conclusions

By integrating all these technologies into a platform which can be managed via a remote web based interface, management of the complete life-cycle of a large scale multi-agent experiment should be substantially facilitated, overcoming the black box perception generated by grid infrastructures. The platform can serve as an important auxiliary tool for ML researchers eager to tap available computational resources, and can be extended to accommodate large-scale experiments for other scientific fields too.

Key future work on the research front revolves around the experimentation with specific games in order to design, schedule and analyze learning/playing experiments, with the expectation that the massive scale of agent interactions will uncover interesting playing behaviors. On the technology front, we expect to fine tune the interface of the platform to a variety of third-party games so that user take-up is facilitated as much as possible.

## Acknowledgment

This work used the European Grid Infrastructure (EGI) through the National Grid Infrastructures NGI_GRNET, HellasGRID as part of the SEE Virtual Organization.

A demo of the platform alongside instructions to integrate it with third-party games can be found at http://www.splendor.gr/platform.

# Appendix
## How to build a new game for an experiment

This section provides some key instructions in how one would go about writing the game logic of a new game, based on the (very simple) RSP template, so that the resulting code could be readily integrated with our MABS platform (using the steps set out in detail in the web site). The RSP template is also available at the web site and provides several package imports and wrapper code. However, the game logic itself (for RSP) is concentrated in just a couple of anchor points, which -in principle- are the only points which one would need to change for a new game.

These anchor points are:

- The (`public int`) method `compareMoves`, which provides the reward scheme at the player level.
- The (`public Move`) method `getMove`, which selects a random move for a computer-controlled player.
- The (`public`) method `RockScissorsPaper`, which generates two computer-controlled players and initializes a series of games between them.

Of course, one can also program new functionality (for example, a minimax player, or a neural-network player), but this would normally be just extra self-contained methods/functions that would be called from within the anchor points.

**Table 1.** Key acnhor points for building new games.

| compareMoves | ```java
public int compareMoves(Move theMove){
    if (this == theMove)
            return 0;

    switch (this) {
        case ROCK:
            return (theMove == SCISSORS ? 1 : -1);
        case PAPER:
            return (theMove == ROCK ? 1 : -1);
        case SCISSORS:
            return (theMove == PAPER ? 1 : -1);
    }
    return 0;
}
``` |
|---|---|

| | |
|---|---|
| **getMove** | ```java
public Move getMove() {
    Move[] moves = Move.values();
    Random random = new Random();
    int index = random.nextInt(moves.length);
    return moves[index];
}
``` |
| **RSP** | ```java
public RockScissorsPaper() {
    whitePlayer = new Computer();
    blackPlayer = new Computer();
}
``` |